\documentclass{article}

\usepackage[eandd, final]{neurips_2026}

\usepackage[utf8]{inputenc} 
\usepackage[T1]{fontenc}    
\usepackage{hyperref}       
\usepackage{url}            
\usepackage{booktabs}       
\usepackage{amsfonts}       
\usepackage{nicefrac}       
\usepackage{microtype}      
\usepackage{xcolor}         
\usepackage{amsmath}
\usepackage{booktabs}
\usepackage{multirow}
\usepackage{graphicx}
\usepackage{xspace}
\usepackage{wrapfig}
 \usepackage{subcaption}
 \usepackage[normalem]{ulem}
 \usepackage{enumitem}
\usepackage{placeins}

\newcommand{\best}[1]{\textbf{#1}}
\newcommand{\second}[1]{\underline{#1}}
\newcommand{\ours}{\textsc{TGH}\xspace}
\newcommand{\oursone}{\textsc{TGH-1}\xspace}
\newcommand{\ourstwo}{\textsc{TGH-2}\xspace}
\newcommand{\bestbase}[1]{\uwave{#1}}

\newcommand{\idlast}{\textsc{ID-Last}\xspace}
\newcommand{\semnn}{\textsc{Sem-NN}\xspace}
\newcommand{\idsem}{\textsc{ID+Sem}\xspace}

\title{An Embarrassingly Simple Graph Heuristic Reveals Shortcut-Solvable Benchmarks for Sequential Recommendation}

\author{
\parbox{\textwidth}{
\centering
Haoyu Han\textsuperscript{1},
Li Ma\textsuperscript{1,}\thanks{Corresponding author},
Hanbing Wang\textsuperscript{1},
Bingheng Li\textsuperscript{1},
Daochen Zha\textsuperscript{2},
Chun How Tan\textsuperscript{2}, \\
Huiji Gao\textsuperscript{2},
Xin Liu\textsuperscript{2},
Stephanie Moyerman\textsuperscript{2},
Sanjeev Katariya\textsuperscript{2},
Hui Liu\textsuperscript{1},
Jiliang Tang\textsuperscript{1} \\
\textsuperscript{1}Michigan State University,
\textsuperscript{2}Airbnb, Inc. \\
\texttt{\{hanhaoy1, mali13, wangh137, libinghe, liuhui7, tangjili\}@msu.edu} \\
\texttt{\{daochen.zha, chunhow.tan, huiji.gao, xin.liu\}@airbnb.com} \\
\texttt{\{stephanie.moyerman, sanjeev.katariya\}@airbnb.com}
}
}

\begin{document}

\maketitle
 
\vspace{-0.2in}
\begin{abstract}
\vspace{-0.1in}
Sequential recommendation is a central task in recommender systems, and recent research has increasingly shifted toward generative recommenders that leverage both sequential patterns and semantic item information. However, these methods are often evaluated on a small set of widely used benchmarks. This raises a natural question: {\it do these benchmarks actually require the advanced modeling capabilities of modern generative recommenders?} We conduct a benchmark audit using an intentionally simple graph heuristic: starting from only the last one or two interacted items, it retrieves candidates from a few-hop item-transition graph and ranks them with item-feature similarity. Surprisingly, despite its simplicity, this heuristic matches or outperforms a broad set of modern baselines on a variety of popular sequential recommendation benchmarks. For example, it achieves relative NDCG@10 improvements of 38.10\% and 44.18\%  over the best competing baseline on the widely used Amazon Review Sports and CDs datasets, respectively.
We further show that this phenomenon is not merely an artifact of a particular heuristic, but reflects shortcut solvability in existing benchmarks. Specifically, we identify three shortcut structures that could make next-item prediction easier than expected: low-branching local transition structure, feature-smooth transitions, and limited dependence on long user histories. These shortcuts need not appear simultaneously. Depending on the dataset, even one or two strong shortcut signals can make simple local retrieval highly competitive, while weakening the relevant signals allows more sophisticated models to show clearer benefits. Our broader evaluation across 14 diverse datasets further shows that model rankings change substantially with dataset properties, while the simple graph heuristic remains competitive on 10 out of 14 datasets. These findings suggest that strong performance on several standard sequential recommendation benchmarks may not faithfully reflect whether recent methods achieve the advanced modeling capabilities they aim to demonstrate. Rather than treating datasets as interchangeable leaderboards, we argue for more careful dataset selection and dataset-level diagnostic analysis when using benchmarks to support claims about the benefits of new recommendation models. Our code is available at \url{https://github.com/haoyuhan1/GraphRec/}.

\end{abstract}

\vspace{-0.2in}
\section{Introduction}
\vspace{-0.1in}
\label{sec:intro}
Sequential recommendation has long been a fundamental problem in recommender systems, where the goal is to infer a user's next interaction from past behaviors~\cite{boka2024survey,pan2026survey,fang2020deep}.  Early studies mainly relied on  collaborative filtering~\cite{hu2008collaborative, ekstrand2011collaborative} and Markov-chain models~\cite{he2017translation}, while later neural methods introduced recurrent networks~\cite{hidasi2015gru4rec,tan2016improved}, graph-based recommenders~\cite{he2020lightgcn,wu2019srgnn,fan2019graph,wang2020traffic}, and attention-based sequence encoders~\cite{zhou2018din,zhou2019dien}. Recently, however, the field has seen a rapid rise of generative recommendation methods~\cite{tiger, wang2024letter, hou2025actionpiece, liu2026beyond}. These generative methods reformulate recommendation as generation or retrieval over discrete item identifiers, semantic IDs, or textual item representations~\cite{geng2022recommendation,hua2023index, liu2026understanding}. As a result, item-side information has become increasingly central: many recent models rely on item text, metadata, semantic embeddings, or LLM-generated representations to construct item codes, prompts, or prediction targets~\cite{wei2025cofirec, hou2025actionpiece, ji2024genrec}.

\begin{wrapfigure}{r}{0.46\linewidth}
    \centering
    \includegraphics[width=\linewidth]{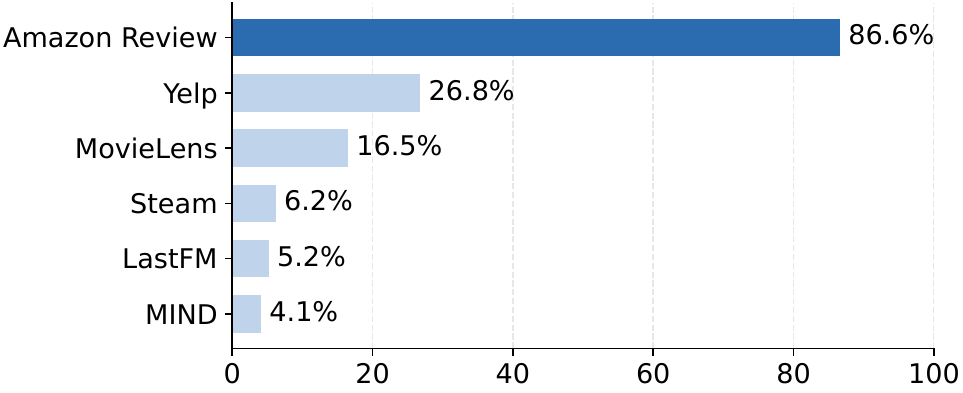}
    \vspace{-8pt}
    \caption{The proportion of surveyed sequential recommendation papers utilizing each dataset.}
    \label{fig:dataset_usage}
    \vspace{-10pt}
\end{wrapfigure}

Along with this methodological shift, evaluation practice has also become strikingly concentrated. 
We analyze dataset usage across 94 recent generative-recommendation papers, detailed in Appendix~\ref{app:sec:papers}. As shown in Figure~\ref{fig:dataset_usage}, Amazon Review datasets~\cite{amazondata} dominate recent evaluations, while other datasets are used much less frequently. This dominance is understandable: Amazon Review datasets provide rich item-side metadata, such as titles, categories, brands, prices, reviews, and other textual fields, which can be naturally processed by text encoders or LLM-based generators~\cite{t5xl, touvron2023llama, han2026connectionmind} for semantic and generative recommendation.

However, this concentration creates a critical question: \emph{what do these dominant benchmarks actually measure?} High performance on these datasets is often taken as evidence that a model better captures user preferences, sequential dynamics, semantic structure, long-range dependencies, or generative reasoning. But this interpretation relies on an implicit assumption: the benchmark truly requires these advanced capabilities. If the same benchmark can be solved by a much simpler signal,  then the comparison may be less informative than it appears. Strong performance may reflect dataset-specific shortcuts rather than genuinely advanced sequential modeling.

In this work, we uncover such shortcuts. We revisit widely used sequential recommendation benchmarks through an intentionally simple graph heuristic. Given only the training sequences, we construct an item-transition graph from observed item-to-item transitions. At test time, the heuristic looks only at the user's last one or two interacted items, retrieves candidates from their few-hop neighborhoods in the transition graph, and ranks them by item-feature similarity with a small direct-edge bonus. It has no deep sequence encoder, no generative objective, no learned user representations, and no training, making it simple and computationally efficient.

Despite its simplicity, this heuristic outperforms a broad set of strong baselines on the Amazon Review benchmarks, including traditional sequential models, graph-based recommenders, transformer-based recommenders, and recent generative recommenders. This result suggests that \textbf{strong performance on these datasets may be achievable through much simpler signals than expected}.
Therefore, relying only on these datasets may not faithfully reflect the advanced modeling capabilities that recent methods aim to demonstrate.
To understand why this happens, we identify three shortcut structures that potentially help explain this phenomenon: low-branching local transition structure, where recent items have small but useful transition neighborhoods; feature-smooth transitions, where transition-connected items are close in feature space; and limited dependence on long user histories, where the last one or two interactions already provide enough signal for next-item prediction.

We then ask whether this phenomenon is unique to Amazon Review benchmarks or reflects a broader issue in sequential recommendation evaluation. To answer this question, we evaluate the same baselines across 14 datasets with diverse transition structures, feature signals, and history dependencies. The simple heuristic remains competitive on 10 of them, suggesting that \textbf{shortcut-solvable structures are widespread in existing benchmarks}.  At the same time, the heuristic is not universally superior: when the relevant shortcut signals are weakened, more sophisticated models can show clearer benefits. These results suggest two cautions for the community. For model designers, we encourage benchmark choices to be aligned with the capabilities being claimed. Methods that claim improved semantic understanding, generative modeling, or long-context reasoning should ideally be evaluated on datasets where simple local-transition, feature-similarity, or short-history shortcuts are insufficient. For dataset and benchmark creators, dataset releases would benefit from accompanying diagnostic analyses of transition structure, feature smoothness, history dependence, and other relevant properties. Such diagnostics can help clarify what a dataset actually measures and reduce the risk that future evaluations mistake shortcut-solvable performance for genuine modeling progress.

\section{Related Work}
\label{sec:related}

\paragraph{Sequential recommendation.}
Sequential recommendation aims to predict the next item a user will interact with given a historical interaction sequence. Early approaches often combine collaborative filtering with sequential transition modeling~\citep{hu2008collaborative}. For example, matrix factorization~\citep{koren2009matrix} learns user and item representations from historical interactions, while Markov-chain-based methods model short-term item-to-item transitions~\cite{rendle2010factorizing}. Later neural methods introduce more expressive sequence encoders to capture user dynamics. GRU4Rec~\cite{hidasi2015gru4rec} uses recurrent neural networks to model session-level behavior, Caser~\cite{tang2018caser} applies convolutional filters over recent interactions, SASRec~\cite{sasrec} uses self-attention to capture item dependencies in user histories, and BERT4Rec~\cite{sun2019bert4rec} adopts bidirectional Transformer-style representation learning with masked item prediction. Beyond sequence encoders, graph-based recommendation methods exploit the structure of user-item or item-item interactions. LightGCN~\cite{he2020lightgcn} performs simplified graph propagation on the user-item interaction graph and serves as a strong collaborative filtering baseline. For sequential and session-based recommendation, SR-GNN~\cite{wu2019srgnn} represents each session as an item-transition graph, while GCE-GNN~\cite{wang2020gcegnn} further combines local session graphs with global item-transition information. These methods show that collaborative structure and item-transition patterns are highly informative for next-item prediction.

More recently, sequential recommendation has moved toward increasingly sophisticated semantic, generative, and LLM-based models~\cite{zhou2020s3,hou2022towards, li2023text}. This trend is partly driven by datasets with rich item-side information~\cite{amazondata, amazon_m2}, where item titles, descriptions, categories, and reviews can be used to construct semantic representations or natural-language prompts. Generative recommenders~\cite{deldjoo2024review, ayemowa2024analysis} reformulate recommendation as item generation or generative retrieval, often by assigning items discrete semantic identifiers or token sequences. HSTU~\cite{zhai2024actions} studies large-scale sequential transduction for recommendation, TIGER~\cite{tiger} formulates recommendation as generative retrieval over semantic item IDs, LETTER~\cite{wang2024letter} integrates semantic information and collaborative signals to learn item tokenizations, and CoFiRec~\cite{wei2025cofirec} learns a coarse-to-fine tokenizer for generative recommendation. LLM-based recommenders~\cite{geng2022recommendation, li2024large} further leverage language models to encode item content, user histories, or recommendation prompts. Although these models are motivated by advanced modeling capabilities such as semantic understanding, generative retrieval, and long-context sequence modeling, many of them are evaluated on a relatively small set of widely used sequential recommendation benchmarks. In contrast to proposing another sophisticated recommender, our work asks whether these benchmarks actually require such advanced modeling capacity.

\paragraph{Evaluation and benchmark auditing.}
Offline evaluation is central to recommender system research, but prior work~\cite{krichene2020sampled, meng2020exploring, zhao2022revisiting} has shown that empirical conclusions can be highly sensitive to preprocessing choices, data splitting, candidate sampling, hyperparameter tuning, and baseline selection. Several studies~\cite{ferrari2019we, ferrari2021troubling, rendle2020neural} further question whether complex neural recommenders always provide genuine progress over strong and well-tuned simpler baselines. Our work follows this benchmark-auditing perspective, but studies a different failure mode: \emph{shortcut solvability} in sequential recommendation benchmarks. Prior critiques mainly examine whether evaluation protocols are fair and whether proposed models are compared against sufficiently strong baselines. In contrast, we ask what signals the benchmarks themselves reward. We show that several widely used sequential recommendation datasets can be largely solved by an embarrassingly simple graph heuristic.

\section{Problem Setup and Diagnostic Heuristic}
\label{sec:problem}
In this section, we introduce the sequential recommendation and our graph-based diagnostic heuristic.

\subsection{Problem Formulation}

Let $\mathcal{U}$ and $\mathcal{I}$ denote the user and item sets, respectively. Each user $u \in \mathcal{U}$ has a chronological interaction sequence
\[
S_u = (i^u_1, i^u_2, \ldots, i^u_{T_u}),
\]
where $i^u_t \in \mathcal{I}$ is the item interacted with at time $t$. Given a prefix sequence
\[
S^u_{1:t} = (i^u_1, \ldots, i^u_t),
\]
the goal of sequential recommendation is to rank candidate items so that the next item $i^u_{t+1}$ appears near the top.

\subsection{A Simple Graph-Heuristic Probe}

We use an intentionally simple graph heuristic as a diagnostic probe. The heuristic is motivated by two basic sources of recommendation signals: collaborative structure from user-item interaction sequences, and semantic similarity from item-side features. Rather than proposing a new recommendation architecture, we use this heuristic to assess how much predictive power these simple signals provide on widely used benchmarks.

Given the training sequences, we construct a directed item transition graph $G=(\mathcal{I}, \mathcal{E})$, where each node is an item $i\in \mathcal{I}$. A directed edge $(i,j) \in \mathcal{E}$ is added if item $j$ appears immediately after item $i$ in a training sequence. Let $N_{i,j}$ denote the number of observed transitions from item $i$ to $j$. We assign each edge a normalized weight
\[
w_{i,j}
=
\frac{\log(1+N_{i,j})}
{\max_{j': N_{i,j'}>0} \log(1+N_{i,j'})},
\]
so that outgoing normalized edge weights from each item are scaled to $[0,1]$.

At inference time, given a prefix sequence $S^u_{1:t}$, the heuristic uses either the last item $i^u_t$ or the last two items $(i^u_{t-1}, i^u_t)$ as anchors. For each anchor item $s$, it retrieves candidates from a few-hop neighborhood of $s$ in the transition graph. Within each hop, candidates are ranked mainly by feature similarity to the anchor, with a small edge-weight bonus for direct $1$-hop neighbors. This bonus reflects the intuition that larger edge weights correspond to more frequent transitions and thus higher transition likelihood.

Specifically, each item $i$ has an L2-normalized feature embedding $\hat{\mathbf{e}}_i$. For a candidate item $c$ retrieved from the $\ell$-hop neighborhood of anchor $s$, we compute
\begin{align}
\label{eq:1}
\mathrm{score}_s(c)
=
\hat{\mathbf{e}}_s^\top \hat{\mathbf{e}}_c
+
\alpha \cdot \mathbf{1}[\ell=1] \cdot w_{s,c},
\end{align}
where $\alpha$ is a small edge-weight bonus weight.
The first term is the cosine similarity between the anchor and the candidate. The second term is a direct-transition bonus based on the normalized edge weight, and is applied only to $1$-hop candidates. For candidates from hops larger than one, the score reduces to feature similarity.

For each anchor and each hop, we keep only a small number of top-scoring candidates. If the same item is retrieved from multiple anchors or hops, we keep its highest score. The final recommendation list is obtained by sorting all retained candidates by their scores.

We refer to this heuristic as \ours, short for Transition-Graph Heuristic. It is deliberately simple: it uses only recent anchor items, local transition-graph neighborhoods, and item-feature similarity, without a sequence encoder, generative objective, representation learning, or training. As a result, it is substantially more efficient than existing learnable sequential recommenders, which typically require extensive training.

\subsection{Experimental Setting}

\paragraph{Evaluation protocol.}
We follow the standard next-item recommendation protocol~\cite{sasrec, tiger}. For each user, interactions are ordered chronologically, and models are evaluated by ranking the held-out next item given a prefix sequence. We report top-$K$ ranking metrics, including Recall@K and NDCG@K, where $K \in [1, 5, 10]$.

\paragraph{Text representation.}
To ensure a fair comparison among methods that use item-side textual information, we use a shared text encoder for all of them. Specifically, we encode item text with \texttt{google/flan-t5-xl}~\cite{t5xl}, following~\cite{snap}. The input text is constructed from the available metadata in each dataset, such as item title, category, description, or other textual fields.

\paragraph{Graph-heuristic variants.}
We evaluate two short-context variants of \ours. Although the retrieval budgets and hyperparameters could be tuned on the validation set, we fix them across all datasets to keep the heuristic deliberately simple and ensure that its performance is not driven by dataset-specific configuration choices.

In the \textbf{\oursone} setting, \ours uses only the last interacted item $i_t^u$ as the anchor. It considers the anchor's $1$-, $2$-, and $3$-hop neighborhoods in the transition graph, and selects the top $(7,2,1)$ candidates from these hops according to the scoring function in Equation~\ref{eq:1}, respectively.

In the \textbf{\ourstwo} setting, \ours uses both the last item $i_t^u$ and the second-last item $i_{t-1}^u$ as anchors. For the last item, it selects the top $(5,1)$ candidates from its $1$- and $2$-hop neighborhoods. For the second-to-last item, it selects the top $(3,1)$ candidates from its $1$- and $2$-hop neighborhoods.

We also set the edge-bonus weight to $\alpha=0.5$ for all datasets.

\paragraph{Compared methods.}
We compare the heuristic with representative methods from multiple recommendation families, including graph-based recommendation methods such as LightGCN~\cite{he2020lightgcn} and SR-GNN~\cite{wu2019srgnn}, conventional sequential recommenders such as SASRec~\cite{sasrec}, and recent generative recommendation methods such as HSTU~\cite{zhai2024actions}, TIGER~\cite{tiger}, LETTER~\cite{wang2024letter}, and CoFiRec~\cite{wei2025cofirec}.

\section{Experiments}
\label{sec:exp}
\subsection{The Graph Heuristic is Surprisingly Strong on Amazon Review Benchmarks}

\begin{table}[t]
\centering
\caption{Main results on Amazon Review benchmarks. We report Recall@10 and NDCG@10. All results are percentages. Best results are in \textbf{bold}, second-best results are \underline{underlined}, and the best baseline is denoted by \uwave{wavy underline}.}
\label{tab:amazon_main}
\resizebox{\linewidth}{!}{
\begin{tabular}{lcccccccc}
\toprule
\multirow{2}{*}{Method}
& \multicolumn{2}{c}{Beauty}
& \multicolumn{2}{c}{Sports}
& \multicolumn{2}{c}{Toys}
& \multicolumn{2}{c}{CDs} \\
\cmidrule(lr){2-3} \cmidrule(lr){4-5} \cmidrule(lr){6-7} \cmidrule(lr){8-9}
& R@10 & N@10 & R@10 & N@10 & R@10 & N@10 & R@10 & N@10 \\
\midrule
LightGCN          & \bestbase{7.21} & \bestbase{4.36} & 3.69 & \bestbase{2.10} & \bestbase{8.63} & \bestbase{5.32} & 2.63 & 1.52 \\
SR-GNN             & 4.98 & 2.82 & 2.61 & 1.33 & 3.77 & 2.32 & 0.63   & 0.35   \\
SASRec            & 6.21 & 3.31 & 3.23 & 1.83 & 7.41 & 4.23 & 4.44 & 2.33 \\
HSTU              & 5.73 & 3.00 & 2.48 & 1.25 & 6.40 & 3.56 & 5.62 & 2.91 \\
TIGER             & 6.41 & 3.60 & 3.70 & 1.96 & 6.01 & 3.27 & 5.06 & 2.78 \\
LETTER            & 6.11 & 3.26 & 3.16 & 1.72 & 5.67 & 3.01 & 5.14 & 2.85 \\
CoFiRec           & 6.24 & 3.36 & \bestbase{3.82} & 2.02 & 5.63 & 2.89 & \bestbase{5.64}   & \bestbase{3.01}   \\
\midrule
\oursone          & \second{7.66} & \second{5.01} & \second{4.27} & \second{2.76} & \second{9.13} & \second{6.12} & \second{6.21} & \second{4.13} \\
\ourstwo          & \best{7.85} & \best{5.07} & \best{4.66} & \best{2.90} & \best{9.44} & \best{6.25} & \best{6.89} & \best{4.34} \\
 Rel. Improv. & +8.88\% & +16.28\% & +21.99\% & +38.10\% & +9.39\% & +17.48\% & +22.16\% & +44.18\% \\
\bottomrule
\end{tabular}
}
\vspace{-0.2in}
\end{table}

We first evaluate on the most widely used Amazon Review benchmarks~\cite{amazondata}, including Beauty, Sports, Toys and CDs. Table~\ref{tab:amazon_main} reports Recall@10 (R@10) and NDCG@10 (N@10); full results with additional metrics are provided in Appendix~\ref{sec:app:results-amazon}. Despite its simplicity, \ours is highly competitive across all four Amazon datasets. In particular, it achieves relative NDCG@10 improvements of 38.10\% and 44.18\% over the best competing baselines on Sports and CDs, respectively. 

These results are surprising because \ours retrieves only a small number of candidates from local transition-graph neighborhoods and ranks them with item-feature similarity, yet still achieves strong performance. This suggests that the Amazon Review benchmarks contain simple predictive signals that can be exploited without advanced sequential or generative modeling. Considering that Amazon Review datasets are used in 86.6\% of the surveyed generative recommendation papers, as shown in Figure~\ref{fig:dataset_usage}, this finding raises concerns about whether relying heavily on performance on these benchmarks provides sufficient evidence for the claimed benefits of modern recommendation models.

\subsection{What Signals Make the Heuristic Work?}

Although \ours achieves surprisingly strong performance on Amazon Review benchmarks, it remains unclear why such a simple heuristic can outperform much more sophisticated models. To understand the underlying signals, we analyze dataset statistics and introduce a set of diagnostic baselines.

\paragraph{Dataset statistics.}
We consider both general dataset statistics and item-transition graph statistics. The general statistics include the number of users, the number of items, and the average sequence length (Avg. Seq. Len.). For the item-transition graph constructed from the training sequences, we report the average out-degree (Avg. Out-Deg.), the average edge weight (Avg. Edge W.), and the coverage of the ground-truth item within the 1-, 2-, and 3-hop neighborhoods of the last interacted item (Cov@1, Cov@2, and Cov@3).

\paragraph{Diagnostic baselines.}
We introduce three controlled baselines to isolate simple signals in the data. 
First, \idlast learns a trainable embedding for each item ID and uses only the last interacted item to predict the next item. This baseline measures how much signal can be captured from the ID-based transitions.
Second, \semnn is a training-free semantic nearest-neighbor baseline. Given the last interacted item, \semnn ranks all candidate items by cosine similarity between their text embeddings and the text embedding of the last item. This baseline tests whether item-feature similarity alone is predictive for next-item retrieval.
Finally, \idsem combines the scores of \idlast and \semnn through late fusion, where the final score is the sum of the ID-based score and the semantic-similarity score. This baseline tests whether ID-based transition signals and item-feature similarity provide complementary information. We further conduct a history-window ablation on SASRec and HSTU. Instead of using the full interaction sequence, we restrict the maximum history window to $1$, so that each prediction can only condition on the immediately previous item. We refer to these variants as \textsc{Last-1}.

\begin{table}[t]
\centering
\caption{Statistics of the item-transition graphs on Amazon Review benchmarks. Coverage@k denotes the percentage of test targets reachable within $k$ hops from the last interacted item.}
\label{tab:amazon-graph-statistics}
\resizebox{\linewidth}{!}{
\begin{tabular}{cccccccccc}
\toprule
Dataset 
& \#Users 
& \#Items
& \#Edges 
& Avg. Seq. Len. 
& Avg. Out-Deg. 
& Avg. Edge W. 
& Cov@1 
& Cov@2 
& Cov@3 \\
\midrule
Beauty & 22,363 & 12,101 & 114,582 & 8.15  & 9.47  & 1.15 & 8.61\% & 24.85\% & 56.64\% \\
Sports & 35,598 & 18,357 & 180,610 & 7.96  & 9.84  & 1.05 & 5.13\% & 20.26\% & 58.16\% \\
Toys   & 19,412 & 11,924 & 102,268 & 7.97  & 8.58  & 1.07 & 8.06\% & 20.16\% & 47.16\% \\
CDs    & 75,258 & 64,443 & 810,347 & 14.58 & 12.57 & 1.08 & 9.07\% & 28.06\% & 61.44\% \\
\bottomrule
\end{tabular}
}
\vspace{-0.1in}
\end{table}

\begin{table}[t]
\centering
\caption{Results of diagnostic baselines on Amazon review benchmarks.}
\label{tab:amazon_diagnostic}
\resizebox{0.7\linewidth}{!}{
\begin{tabular}{@{}lcccccccc@{}}
\toprule
\multirow{2}{*}{Method}
& \multicolumn{2}{c}{Beauty}
& \multicolumn{2}{c}{Sports}
& \multicolumn{2}{c}{Toys}
& \multicolumn{2}{c}{CDs} \\
\cmidrule(lr){2-3} \cmidrule(lr){4-5} \cmidrule(lr){6-7} \cmidrule(l){8-9}
& R@10 & N@10 & R@10 & N@10 & R@10 & N@10 & R@10 & N@10 \\
\midrule
\idlast & 6.16 & 3.74 & 2.70 & 1.63 & 6.81 & 4.22 & 4.91 & 3.00 \\
\semnn  & 5.32 & 3.30 & 2.68 & 1.55 & 8.04 & 5.01 & 1.20 & 0.75 \\
\idsem  & 6.66 & 4.08 & 3.18 & 1.91 & 8.53 & 5.37 & 5.02 & 3.05 \\
\midrule
SASRec  & 6.21 & 3.31 & 3.23 & 1.83 & 7.41 & 4.23 & 4.44 & 2.33 \\
\quad -Last-1 & 5.98 & 3.33 & 3.05 & 1.66 & 6.59 & 3.89 & 3.77 & 1.94 \\
HSTU & 5.73 & 3.00 & 2.48 & 1.25 & 6.40 & 3.56 & 5.62 & 2.91 \\
\quad -Last-1 & 5.39 & 2.92 & 2.61 & 1.37 & 6.20 & 3.42 & 4.38 & 2.26 \\
\midrule
\oursone & \textbf{7.66} & \textbf{5.01} 
         & \textbf{4.27} & \textbf{2.76} 
         & \textbf{9.13} & \textbf{6.12} 
         & \textbf{6.21} & \textbf{4.13} \\
\bottomrule
\end{tabular}
}
\vspace{-0.1in}
\end{table}

The item-transition graph statistics are shown in Table~\ref{tab:amazon-graph-statistics}, and the performance of the diagnostic baselines is shown in Table~\ref{tab:amazon_diagnostic}. These results lead to the following key findings.

\paragraph{Finding 1: many targets are reachable through low-branching local transitions.}
As shown in Table~\ref{tab:amazon-graph-statistics}, the Amazon Review transition graphs have relatively small average out-degrees compared with the total number of items. For example, the average out-degree is only $9.47$ for Beauty and $12.57$ for CDs, while the datasets contain thousands to tens of thousands of items. Meanwhile, the direct $1$-hop neighbors already cover a non-trivial fraction of test targets, with Cov@1 of $8.61\%$ and $9.07\%$ on Beauty and CDs, respectively. This indicates that recent items provide a small but informative local candidate space. After expanding to a few hops, the target coverage increases substantially while the search remains restricted to local transition neighborhoods. We refer to this shortcut structure as \emph{low-branching local transition structure}: the transition graph narrows next-item prediction from ranking over the full item universe to selecting among a small set of locally plausible candidates.

\paragraph{Finding 2: item features provide a useful ranking signal.}
As shown in Table~\ref{tab:amazon_diagnostic}, \semnn retrieves from the full item set using only feature similarity, yet already achieves strong performance on Beauty, Sports, and Toys datasets. This suggests that items adjacent in user sequences often have similar item features on these datasets, making feature similarity a useful signal for next-item prediction. We refer to this shortcut structure as \emph{feature-smooth transitions}.
However, this property does not hold equally across all Amazon Review datasets. On CDs, \semnn performs much worse, indicating that global feature similarity alone may be insufficient.

\paragraph{Finding 3: many Amazon Review benchmarks require little long-context information.}
The strong performance of \oursone suggests that the most recent interaction already contains much of the useful signal. This observation is also supported by \idlast, which uses only the last item ID to predict the next item and still achieves strong performance on several Amazon Review datasets.
To further examine whether long-range history is necessary, we conduct a history-window ablation on SASRec and HSTU. Specifically, we compare the full-history models with their \textsc{Last-1} variants, where each prediction only conditions on the immediately previous item. As shown in Table~\ref{tab:amazon_diagnostic}, using only a history window of size $1$ achieves performance close to the full-history setting on Beauty, Sports, and Toys. This suggests that long-range sequential information may not be essential for strong performance on these benchmarks.
We refer to this shortcut structure as \emph{limited dependence on long user histories}: the benchmark can be largely solved using very recent interactions, without requiring models to capture long-range user preference or complex sequential dynamics.

Overall, these analyses suggest that the three shortcut structures could serve as useful diagnostic axes for interpreting the strong performance of \ours on Amazon Review datasets. While they are not meant to be an exhaustive explanation of benchmark behavior, they help characterize several simple signals that can make a dataset shortcut-solvable. The heuristic can benefit from low-branching local transition structure, feature-smooth transitions, and limited dependence on long user histories, but these shortcuts need not appear simultaneously. For example, CDs shows a weak global feature-similarity signal, as \semnn performs poorly when retrieving from the full item set. Yet \ours remains effective, suggesting that the transition graph first restricts retrieval to a local candidate set, where feature similarity becomes more useful for ranking. This example illustrates that shortcut-solvability may arise from the interaction of multiple simple signals, rather than requiring all identified shortcut structures to be present at once.

\subsection{Beyond Amazon Review: Shortcut Solvability Across Diverse Benchmarks}

The previous subsection shows that \ours performs surprisingly well on Amazon Review benchmarks and identifies three potential dataset shortcuts that could help explain this behavior. We next ask whether this phenomenon is specific to Amazon Review datasets, or whether similar shortcut-solvable patterns also appear in other sequential recommendation benchmarks with item-side information. To this end, we evaluate on a broader collection of datasets, including Delicious~\cite{hetrec2011}, LastFM~\cite{hetrec2011}, MovieLens-1M (ML-1M)~\cite{harper2015movielens}, Yelp~\cite{yelp}, MIND~\cite{wu2020mind}, Goodreads-Comics (GR-Comics)~\cite{goodreads}, Goodreads-Children (GR-Children)~\cite{goodreads}, STEAM~\cite{sasrec}, H\&M~\cite{hm}, and Amazon-M2-UK (Amazon-UK)~\cite{amazon_m2}. The dataset statistics are shown in Table~\ref{tab:graph-statistics}. Compared with the Amazon Review datasets, these benchmarks cover more diverse data regimes, with substantially different sequence lengths, transition-graph densities, and average out-degrees. For example, some datasets have much longer user histories, while others have much larger local transition neighborhoods. This diversity allows us to examine whether the proposed shortcuts help interpret model behavior beyond the Amazon Review setting. We use the same baseline suite as in the Amazon Review experiments, except for CoFiRec, which is specifically designed for Amazon Review datasets.

Table~\ref{tab:other-ndcg10} reports NDCG@10 on the broader set of datasets; additional metrics are provided in Appendix~\ref{sec:app:results-others}.  Overall, \ours remains competitive beyond Amazon Review benchmarks, achieving the best or second-best performance on 6 out of 10 datasets and comparable performance on Amazon-M2-UK. However, \ours is not universally superior. On MovieLens-1M, Yelp, and MIND, it underperforms traditional and generative sequential recommenders. We next use the shortcut structures identified from the Amazon Review analysis to interpret both the success and failure cases of \ours across these broader datasets.

\paragraph{Shortcut 1: low-branching local transition structure.}
As shown in Table~\ref{tab:graph-statistics}, Delicious, LastFM, and Amazon-M2-UK have average out-degrees below $10$, indicating low-branching local transition graphs. On these datasets, the local neighborhoods of recent items provide compact candidate spaces, and \ours achieves strong performance. In contrast, MovieLens-1M has a much larger average out-degree of $78.71$. Although many targets may still be reachable within a few hops, the local candidate space is much larger, making fixed-budget local retrieval less effective. This helps explain why \ours underperforms stronger sequential recommenders on MovieLens-1M.

\paragraph{Shortcut 2: feature-smooth transitions.}
Some datasets remain favorable to \ours even when their transition graphs are not low-branching. For example, Steam and H\&M have large average out-degrees of $116.81$ and $186.54$, respectively, yet \ours still performs well. We attribute this to feature-smooth transitions: adjacent items in user sequences tend to have similar item features. This is supported by the strong performance of \semnn, which retrieves items using only semantic similarity.
In contrast, Yelp has a much smaller average out-degree of $10.96$, but \ours still underperforms the baselines. This suggests that low branching alone is not sufficient. On Yelp, adjacent items are less feature-smooth, as indicated by the weak performance of \semnn. Therefore, even though the local candidate space is relatively small, feature similarity cannot reliably rank the correct next item.

\begin{wrapfigure}{r}{0.5\linewidth}
    \centering
    \includegraphics[width=\linewidth]{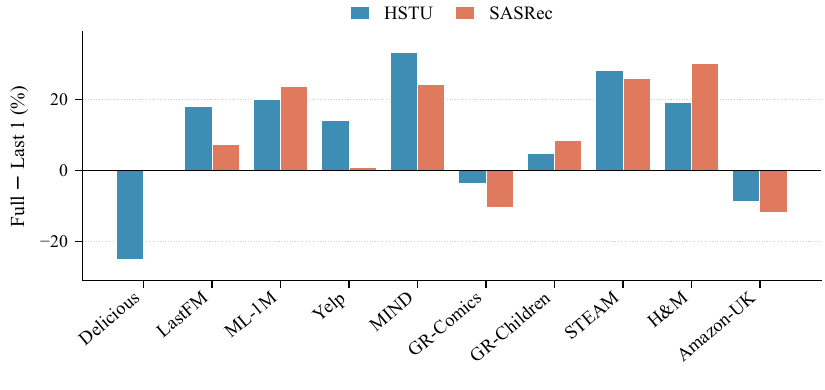}
    \vspace{-8pt}
    \caption{The relative performance gap between the Full sequence and Last-1 settings.}
    \label{fig:performance_gap}
    \vspace{-10pt}
\end{wrapfigure}

\paragraph{Shortcut 3: limited dependence on long user histories.}
To test whether each dataset requires long-range user-history information, we compare SASRec and HSTU under two settings: the full-sequence setting and the \textsc{Last-1} setting, where each prediction only uses the immediately previous item. Figure~\ref{fig:performance_gap} reports the relative performance gap between these two settings. On MovieLens-1M and MIND, the full-sequence models achieve much stronger performance than their \textsc{Last-1} variants, suggesting that these datasets rely more heavily on long-range user histories. Since \ours only uses the last one or two items as anchors, it cannot capture this long-context signal, which helps explain its weaker performance on these datasets.

\begin{table}[t]
\centering
\caption{Statistics of the item-transition graphs on the ten additional benchmarks. Coverage@k denotes the percentage of test targets reachable within $k$ hops from the last interacted item.}
\label{tab:graph-statistics}
\resizebox{\linewidth}{!}{
\begin{tabular}{cccccccccc}
\toprule
Dataset 
& \#Users 
& \#Items
& \#Edges 
& Avg. Seq. Len. 
& Avg. Out-Deg. 
& Avg. Edge W. 
& Cov@1 
& Cov@2 
& Cov@3 \\
\midrule
Delicious & 718 & 1,200 & 4,016 & 9.13 & 3.35 & 1.1 & 9.33 & 11.98 & 18.38 \\ 
LastFM & 1,090 & 3,646 & 30,372 & 34.02 & 8.33 & 1.11 & 5.69 & 21.93 & 53.58 \\ 
MovieLens-1M & 6,040 & 3,416 & 268,867 & 74.06 & 78.71 & 1.6 & 48.11 & 96.59 & 99.93 \\ 
Yelp & 30,431 & 20,033 & 219,632 & 10.4 & 10.96 & 1.02 & 5.75 & 27.86 & 73.57 \\ 
MIND & 48,577 & 39,757 & 824,397 & 28.16 & 20.74 & 1.48 & 40.13 & 89.19 & 96.73 \\ 
Goodreads-Comics & 89,186 & 48,623 & 1,282,693 & 33.78 & 26.38 & 2.14 & 49.67 & 84.36 & 97.25 \\ 
Goodreads-Children & 163,143 & 55,221 & 1,622,817 & 24.26 & 29.39 & 2.14 & 57.04 & 88.57 & 98.13 \\
STEAM & 334,728 & 13,047 & 1,524,022 & 12.59 & 116.81 & 2.11 & 59.44 & 97.97 & 99.8 \\ 
H\&M & 1,077,045 & 104,468 & 19,487,762 & 26.01 & 186.54 & 1.27 & 34.32 & 95.28 & 99.72 \\
Amazon-M2-UK & 1,182,181 & 494,409 & 1,500,196 & 5.12 & 3.03 & 1.67 & 30.35 & 43.69 & 52.94 \\ 
\bottomrule
\end{tabular}
}
\end{table}

 \begin{table*}[t]
 \centering
 \caption{NDCG@10 (\%) across datasets. \textbf{Bold}: best per dataset; \underline{underlined}: second-best.}
 \label{tab:other-ndcg10}
 \resizebox{\linewidth}{!}{
 \begin{tabular}{l c c c c c c c c c c}
 \toprule
\textbf{Method} & \textbf{Delicious} & \textbf{LastFM} & \textbf{ML-1M} & \textbf{Yelp} & \textbf{MIND} & \textbf{GR-Comics} & \textbf{GR-Children} & \textbf{STEAM} & \textbf{H\&M} & \textbf{Amazon-UK} \\
 \midrule
\idlast           & 6.64 & 2.11 & 6.86  & 0.94 & 5.72  & 18.70 & 9.39  & 13.40 & 0.79 & 25.38 \\
\semnn     & 2.62 & 2.66 & 2.99  & 0.03 & 0.14  & 8.30  & 1.91  & 13.40 & 3.33 & 21.76 \\
\idsem      & 6.68 & 2.60 & 7.09  & 0.65 & 5.46  & 18.43 & 7.00  & 13.59 & 3.77   & \underline{26.16} \\
 \midrule
 LightGCN          & 4.37 & 2.22 & 4.32  & 0.40 & 2.78  & 19.62 & 8.03  & 1.98  & 2.93 & \textbf{28.78} \\
 SR-GNN            & 4.78 & 1.56 & 8.48  & \underline{1.48} & \textbf{13.87} & 9.26 & 9.37 & 14.93 & 3.54 & 9.12 \\
 SASRec            & \textbf{7.57} & \textbf{3.32} & \textbf{14.21} & 1.33 & \underline{12.67} & 16.08 & 12.10 & 4.49 & 3.68 & 14.10 \\
 HSTU              & 4.35 & 2.01 & 12.15 & 1.08 & 7.49 & 16.57 & 11.50 & 4.32 & 3.45 & 19.10 \\
 TIGER             & 5.09 & 1.45 & 9.98  & 1.05 & 9.79 & 16.02 & 10.48 & 14.24 & 4.10 & 8.10 \\
LETTER            & 2.47 & 1.97 & \underline{13.30} & \textbf{1.54} & 10.46 & \textbf{24.64} & \textbf{16.14} & \textbf{16.42} & 6.17 & 7.78 \\
 \midrule
\oursone  & 7.06 & \underline{3.07} & 9.25 & 0.90 & 5.51 & 23.12 & 12.23 & 14.85 & \underline{8.06} & 24.87 \\
 \ourstwo & \underline{7.54} & 2.87 & 9.39 & 0.98 & 5.32 & \underline{23.66} & \underline{12.46} & \underline{14.95} & \textbf{8.70} & 25.13 \\
 \bottomrule
 \end{tabular}
 }
 \end{table*}

Overall, across the four Amazon Review benchmarks and ten additional datasets, \ours achieves competitive performance, ranking best or second-best on 10 out of 14 datasets. Its successes and failures can be potentially interpreted through the three shortcut structures identified above. Importantly, this does not imply that modern sequential or generative recommenders are ineffective. Rather, it shows that some benchmarks allow simple shortcuts to dominate evaluation, while other datasets expose prediction behaviors that \ours cannot capture. We therefore next examine prediction-level differences between \ours and learned recommenders to better understand what advanced models may capture beyond the heuristic.

\subsection{Prediction-Level Differences Between the Heuristic and Learned Models}

\begin{figure}[htb]
  \centering
  \begin{subfigure}{0.49\linewidth}
    \includegraphics[width=\linewidth]{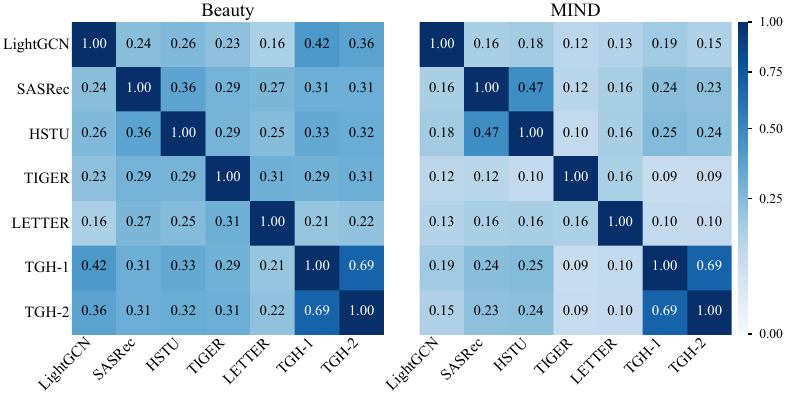}
    \caption{Prediction overlap analysis.}
  \end{subfigure}\hfill
  \begin{subfigure}{0.49\linewidth}
    \includegraphics[width=\linewidth]{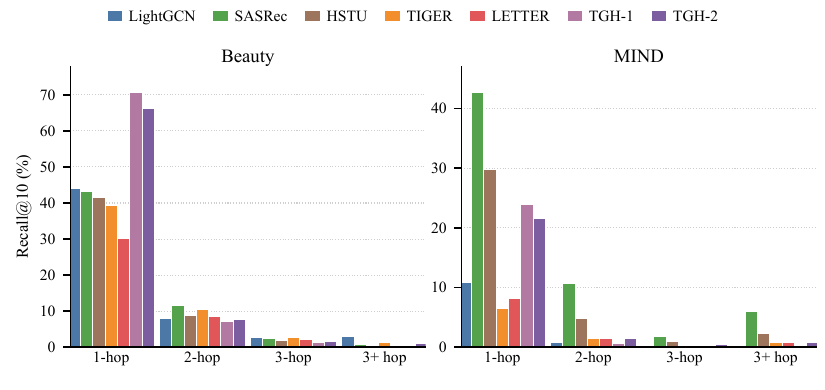}
    \caption{Recall@10 by ground-truth hop distance.}
  \end{subfigure}
  \caption{Prediction-level comparison between \ours and learned recommenders.}
  \label{fig:prediction_difference}
  \vspace{-0.2in}
\end{figure}

The previous results show that \ours can be highly competitive, but this does not imply that learned sequential or generative recommenders are ineffective. To better understand the difference between \ours and learned models, we further analyze their prediction patterns.

We first measure the Jaccard similarity between the correct-prediction sets of different methods. Figure~\ref{fig:prediction_difference}(a) shows the overlap analysis on Beauty and MIND. The overlap between \ours and learned sequential recommenders is relatively low, indicating that they often succeed on different test instances. This suggests that, although \ours can achieve competitive overall performance by exploiting benchmark shortcuts, learned models capture additional signals that are not fully covered by the heuristic.

We further evaluate Recall@10 grouped by the hop distance between the last interacted item and the ground-truth item in the transition graph. As shown in Figure~\ref{fig:prediction_difference}(b), \ours performs best when the ground-truth item is within a small number of hops, which is expected given its local retrieval mechanism. In contrast, learned sequential recommenders achieve stronger performance on harder cases, especially when the ground-truth item is beyond $3$ hops. This indicates that learned models can capture prediction patterns beyond simple local retrieval, such as longer-range sequential dependencies or more complex user-behavior signals.

Overall, this analysis shows that the issue is not that current baselines are weak. Rather, many widely used benchmarks contain shortcuts that can be solved by simple heuristics, so aggregating the overall performance on these datasets alone may be unreliable evidence for the advantages of new methods. For model designers, this calls for evaluation on datasets that truly test the claimed capability, such as settings where local-transition retrieval, feature similarity, or short-history prediction is insufficient. For benchmark designers, this calls for reporting dataset-level diagnostics, such as transition branching, feature-smoothness, and history dependence, to make clear what signals the benchmark rewards.

\subsection{Evaluation under a Global Temporal Split}
\label{sec:global-temporal-split}

Our main experiments use leave-one-out (LOO) evaluation, the most common evaluation protocol in prior sequential recommendation studies~\cite{tiger, wang2024letter, wei2025cofirec}. We additionally evaluate \ours under a global temporal split (GTS)~\cite{gusak2025time, hidasi2023widespread}, where interactions are divided according to a shared chronological boundary across users. Following \cite{gusak2025time}, the test cutoff $T_{\text{test}}$ is the 0.9
quantile of all interaction timestamps, and the validation cutoff
$T_{\text{val}}$ is the 0.9 quantile of the timestamps before $T_{\text{test}}$. All models are trained only on interactions before $T_{\text{val}}$. For validation, the target is each user's last interaction in
$[T_{\text{val}}, T_{\text{test}})$, given all earlier interactions.
For testing, we use the \emph{successive} protocol: every interaction at or
after $T_{\text{test}}$ is predicted in turn from all interactions that precede it. Metrics are averaged first over each user's targets and then over users.

As shown in Table~\ref{tab:gts-results}, \ours achieves the highest Recall@10 on five of the six datasets: Beauty, Sports, CDs, Delicious, and GR-Comics. On ML-1M, learned sequential and generative recommenders perform better. These results are consistent with our findings under the LOO setting, showing that \ours remains competitive under GTS when evaluation follows a realistic chronological boundary.

\begin{table}[!htb]
\centering
\caption{Recall@10 (\%) under a global temporal split. Best results per dataset are in \textbf{bold}.}
\label{tab:gts-results}
\small
\begin{tabular}{lrrrrrr}
\toprule
Method & Beauty & Sports & CDs & Delicious & ML-1M & GR-Comics \\
\midrule
SASRec & 2.07 & 1.80 & 0.71 & 8.80 & 14.23 & 12.63 \\
SR-GNN & 2.23 & 1.71 & 0.46 & 2.00 & 10.79 & 7.34 \\
HSTU & 1.98 & 1.87 & 1.03 & 3.19 & 13.51 & 15.78 \\
TIGER & 1.79 & 1.66 & 0.61 & 8.60 & 12.42 & 16.41 \\
LETTER & 2.54 & 2.48 & 1.52 & 4.09 & \textbf{14.38} & 22.86 \\
\midrule
\oursone & 3.28 & 2.97 & 2.24 & 8.52 & 10.09 & 24.16 \\
\ourstwo & \textbf{3.34} & \textbf{3.08} & \textbf{2.37} & \textbf{10.46} & 10.62 & \textbf{26.06} \\
\bottomrule
\end{tabular}
\end{table}

\subsection{Sensitivity to the Text Encoder}
\label{sec:text-encoder-sensitivity}

To examine sensitivity to the item text representation, we evaluate a different text encoder, GTR-T5-XL~\cite{ni2022large}, in place of the original FLAN-T5-XL~\cite{t5xl}. Table~\ref{tab:text-encoder-results} reports results on Beauty and Sports for \ours and three learned baselines. Although individual scores change with the encoder, both \oursone and \ourstwo outperform the compared learned baselines on Recall@10 and NDCG@10 with either encoder. Thus, the advantage of the heuristic is not specific to a fixed text encoder.

\begin{table}[!htb]
\centering
\caption{Text-encoder comparison on Beauty and Sports. Each entry reports GTR-T5-XL, with the original FLAN-T5-XL result in parentheses. R@10 and N@10 denote Recall@10 and NDCG@10 (\%). Best results for each encoder and metric are in \textbf{bold}.}
\label{tab:text-encoder-results}
\small
\begin{tabular}{lrrrr}
\toprule
\multirow{2}{*}{Method} & \multicolumn{2}{c}{Beauty} & \multicolumn{2}{c}{Sports} \\
\cmidrule(lr){2-3}\cmidrule(lr){4-5}
& R@10 & N@10 & R@10 & N@10 \\
\midrule
SR-GNN & 5.30 (4.98) & 2.79 (2.82) & 2.72 (2.61) & 1.46 (1.33) \\
TIGER & 6.68 (6.41) & 3.65 (3.60) & 3.66 (3.70) & 1.99 (1.96) \\
LETTER & 6.20 (6.11) & 3.32 (3.26) & 3.42 (3.16) & 1.81 (1.72) \\
\midrule
\oursone & 8.03 (7.66) & 5.18 (5.01) & 4.54 (4.27) & 2.88 (2.76) \\
\ourstwo & \textbf{8.16 (7.85)} & \textbf{5.23 (5.07)} & \textbf{4.87 (4.66)} & \textbf{2.99 (2.90)} \\
\bottomrule
\end{tabular}
\end{table}

\vspace{-0.1in}
\section{Conclusion}
\label{sec:conclusion}
In this work, we revisit the evaluation practice of sequential recommendation, especially in the context of recent generative recommenders. We show that an intentionally simple transition-graph heuristic can achieve competitive performance on many widely used benchmarks, including several Amazon Review datasets that dominate recent evaluations. This result suggests that strong benchmark performance may sometimes be driven by simple shortcut signals rather than advanced sequential or generative modeling ability.
Through dataset statistics, diagnostic baselines, and cross-dataset evaluation, we identify three shortcut structures that help explain this behavior: low-branching local transition structure, feature-smooth transitions, and limited dependence on long user histories. Across 14 datasets, the heuristic is competitive on 10, but it is not universally superior; when these shortcut signals are weakened, learned sequential and generative models can show clearer benefits.

Our findings highlight the need for more capability-aware evaluation. We encourage model designers to select datasets that directly test the capabilities their methods aim to improve, and benchmark designers to report dataset-level diagnostics that clarify what signals a benchmark rewards. Rather than treating benchmarks as interchangeable leaderboards, future work can use them as tools for understanding when and why different recommendation models succeed.

\section*{Acknowledgments}

Haoyu Han, Hanbing Wang, Bingheng Li, Hui Liu and Jiliang Tang are supported by the National Science Foundation (NSF) under grant numbers IIS2212032, IIS2212144, IIS 2504089, DUE2234015, CNS2246050, and DRL2405483, the Michigan Department of Agriculture and Rural Development, the U.S. Department of Energy, Office of Science, Office of Advanced Scientific Computing Research, under Award Number DE-SC0026773.

\bibliographystyle{unsrt}
\bibliography{reference}


\appendix
\newpage

\section{Details of Surveyed Papers}
\label{app:sec:papers}

\begin{figure}[htb]
    \centering
    \begin{subfigure}[t]{0.48\linewidth}
      \centering
      \includegraphics[width=\linewidth]{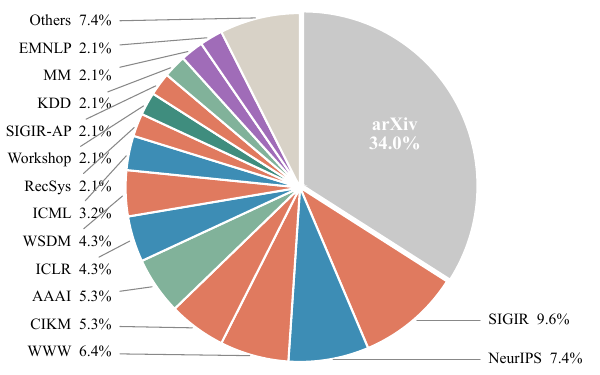}
      \caption{Distribution of publication venues.}
      \label{fig:venue-pie}
    \end{subfigure}\hfill
    \begin{subfigure}[t]{0.48\linewidth}
      \centering
      \includegraphics[width=\linewidth]{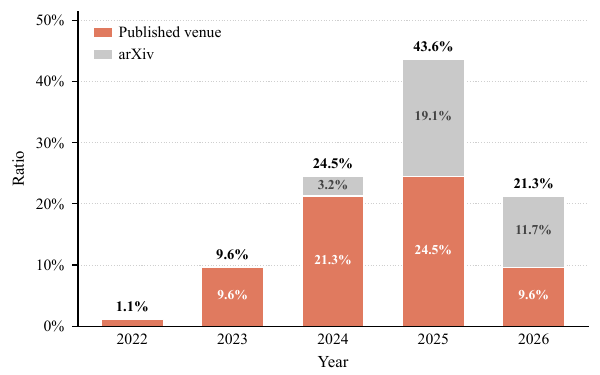}
      \caption{Yearly share (venue vs.\ arXiv).}
      \label{fig:year-bar}
    \end{subfigure}
    \caption{Statistics of the surveyed papers.}
    \label{fig:corpus-stats}
  \end{figure}

We collect 94 papers published between 2022 and 2026 that propose or evaluate \emph{generative recommendation} methods; the full list of paper titles is provided in our code repository. The corpus statistics are summarized in Figure~\ref{fig:corpus-stats}. These papers are gathered from major recommendation, information retrieval, machine learning, and data mining venues, as well as from arXiv. As shown in Figure~\ref{fig:venue-pie}, arXiv preprints account for approximately $34.0\%$ of the corpus, while the remaining papers are published in venues such as SIGIR, NeurIPS, WWW, CIKM, AAAI, ICLR, WSDM, ICML, and RecSys. Figure~\ref{fig:year-bar} further shows the rapid growth of generative recommendation research in recent years. The field expanded sharply after 2023: papers from 2024 account for $24.5\%$ of the corpus, papers from 2025 account for $43.6\%$, and papers from 2026 already account for $21.3\%$, including preprints and accepted-but-not-yet-published works.

\section{Datasets}
\label{app:sec:datasets}

We conduct experiments on 14 widely used recommendation datasets covering e-commerce, review, news, music, movie, and game recommendation scenarios. To ensure fair comparisons with prior work, each dataset is preprocessed following the protocols used in previous studies. In general, we construct user interaction sequences according to interaction timestamps.

\textbf{Amazon Review benchmarks}~\cite{amazondata}. 
The Amazon Review benchmarks are category-specific product recommendation datasets constructed from user reviews and interaction histories. 
We use four subsets: \textit{Beauty}, \textit{Sports}, \textit{Toys}, and \textit{CDs}. 
Following TIGER~\cite{tiger}, GRID~\cite{snap}, and ActionPiece~\cite{hou2025actionpiece}, we filter out users with fewer than 5 interactions. 
For item textual information, we retain the \textit{title}, \textit{price}, \textit{brand}, and \textit{categories} fields.

\textbf{Delicious}~\cite{hetrec2011}. 
Delicious is a social bookmarking dataset that captures user interactions with web resources and their associated tags. 
We filter out users and items with fewer than 5 interactions and retain the \textit{title} and \textit{tags} as textual fields.

\textbf{LastFM}~\cite{hetrec2011} and \textbf{MovieLens-1M (ML-1M)}~\cite{harper2015movielens}. 
LastFM is a music recommendation dataset based on user--artist interactions, while ML-1M is a movie recommendation dataset based on user ratings. 
We follow the preprocessing settings used in TokenRec~\cite{qu2025tokenrec}. 
Specifically, we filter out users with fewer than 5 interactions and truncate the maximum sequence length to 100. 
We keep \textit{artist} and \textit{tags} for LastFM, and \textit{title} and \textit{genres} for ML-1M.

\textbf{Yelp}~\cite{yelp}. 
Yelp is a business review dataset that reflects users' preferences over local businesses such as restaurants and services. 
Following LETTER~\cite{wang2024letter}, we filter out users with fewer than 5 interactions and retain the \textit{title} and \textit{description} fields as item textual information.

\textbf{MIND}~\cite{wu2020mind}. 
MIND is a news recommendation dataset built from user click histories over online news articles. 
We use the validation split of the MIND-small dataset as our experimental dataset, which contains approximately 50K users and six weeks of click histories. 
Following~\cite{wu2020mind}, we retain the \textit{news title} as the item textual feature and filter out users with fewer than 3 interactions. 

\textbf{GoodReads}~\cite{goodreads}. 
GoodReads is a book recommendation dataset constructed from users' reading and review histories. 
We use two subsets~\cite{goodreads_small}: \textit{GoodReads-Children} and \textit{GoodReads-Comics}. 
We filter out users with fewer than 3 interactions and retain the \textit{title}, \textit{authors}, \textit{publisher}, \textit{year}, and \textit{description} fields as textual information.

\textbf{Steam}~\cite{sasrec}. 
Steam is a video game recommendation dataset based on user interactions with games on the Steam platform. 
We follow the commonly used preprocessing strategy in~\cite{sasrec} and filter out users with fewer than 3 interactions. 
We retain the \textit{title}, \textit{developer}, \textit{publisher}, and \textit{tags} fields as textual information.

\textbf{Amazon-M2-UK}~\cite{amazon_m2}. 
Amazon-M2 is a multi-market e-commerce dataset designed for recommendation across different regional Amazon marketplaces. 
We use the UK locale from the Amazon-M2 dataset and filter out users with fewer than 3 interactions. 
We retain the \textit{title}, \textit{brand}, \textit{categories}, and \textit{description} fields as textual information.

\section{Baselines}
\label{app:sec:baselines}

We compare our method with the following representative baselines:

\begin{itemize}[leftmargin=*]
    \item \textbf{\idlast} is a pure item-to-item collaborative filtering method trained with the BPR loss. It predicts the next item based on the last interacted item.

    \item \textbf{\semnn} ranks candidate items according to their pretrained language-model embedding similarity to the anchor item, serving as a content-based lower bound.

    \item \textbf{\idsem} combines \idlast and \semnn by integrating collaborative item-transition signals with semantic item similarity.

    \item \textbf{LightGCN}~\cite{he2020lightgcn} simplifies GCN-based collaborative filtering by retaining only linear neighborhood aggregation and using layer-averaged embeddings as the final representation. For fair comparison with sequential baselines, we apply the same propagation rule on a global item-item transition graph constructed from training sequences. We also leverage the item text embeddings as the node features.

    \item \textbf{SR-GNN}~\cite{wu2019srgnn} models each session as a directed graph and uses a gated graph neural network to capture complex item transitions. It then combines global session preference and current interest with attention for next-item recommendation.

    \item \textbf{SASRec}~\cite{sasrec} is a Transformer-based sequential recommender that models user interaction histories with causal self-attention and predicts the next item by matching the sequence representation with item embeddings.

    \item \textbf{HSTU}~\cite{zhai2024actions} is a generative sequential recommender that reformulates ranking and retrieval as sequential transduction tasks. It encodes user actions into a unified sequence and uses efficient hierarchical sequential transduction units for long-sequence recommendation modeling.

    \item \textbf{TIGER}~\cite{tiger} is a generative retrieval framework for sequential recommendation. It represents each item with a short discrete semantic ID produced by an RQ-VAE~\cite{lee2022autoregressive} tokenizer trained on item content embeddings, and trains an encoder--decoder Transformer to autoregressively generate the semantic ID of the next item.

    \item \textbf{LETTER}~\cite{wang2024letter} improves the RQ-VAE tokenizer by introducing collaborative-filtering alignment with SASRec~\cite{sasrec} embeddings and a codebook-diversity loss to alleviate codebook collapse and semantic ID collisions.

    \item \textbf{CoFiRec}~\cite{wei2025cofirec} extends the RQ-VAE tokenizer with a hierarchical design that separately quantizes textual views and collaborative signals into semantic IDs, which are then used by a TIGER-style encoder--decoder generator.
\end{itemize}

\section{Implementation Details}
\label{app:sec:implementation_details}

Following the standard leave-one-out evaluation protocol~\cite{sasrec,tiger}, we use the last interacted item of each user for testing, the second-to-last item for validation, and the remaining interactions for training. For all baselines, we carefully follow the implementation details and hyperparameter settings suggested in their original papers. We truncate each user history to the most recent $50$ items~\cite{hou2025actionpiece} for ID-based methods and $20$ items~\cite{wang2024letter, zhai2024actions} for generative methods. We set the latent dimension to $64$ for all methods, except for LETTER, where we use a latent dimension of $32$ following its original setting.

For generative methods, we reproduce TIGER~\cite{tiger} under the GRID~\cite{snap} architecture, and implement LETTER~\cite{wang2024letter} and CoFiRec~\cite{wei2025cofirec} following their original papers. These methods generally follow a two-stage training recipe. In Stage~1, an RQ-VAE tokenizer compresses each item's pretrained text embedding into a 4-tuple of discrete codes, using $L{=}4$ residual codebooks with codebook size $K{=}256$. In Stage~2, a lightweight T5 encoder--decoder model is trained \emph{from scratch} on user histories of up to $H{=}20$ items, where each item is represented as a flattened sequence of discrete codes. During inference, top-$K$ recommendations are generated by beam search with beam size $10$. Unless otherwise specified, all hyperparameters are kept consistent with those reported in the original papers. We use a single H200 GPU to train and inference all baselines.

\section{Detailed Results on Amazon Review Datasets}
\label{sec:app:results-amazon}

In Section~\ref{sec:exp}, we report only Recall@10 (R@10) and NDCG@10 (N@10) on the four Amazon Review datasets. In this section, we provide the full results for Recall@1, Recall@5, Recall@10, NDCG@1, NDCG@5, and NDCG@10 on Beauty, Sports, Toys, and CDs. The detailed results are shown in Tables~\ref{tab:per-beauty}, \ref{tab:per-sports}, \ref{tab:per-toys}, and \ref{tab:per-cds}, respectively. From the results, we observe that the proposed \ours consistently outperforms all baselines across all datasets and evaluation metrics, further demonstrating its effectiveness on the Amazon Review benchmarks.

 \begin{table}[htb]
 \centering
 \caption{Detailed results on \textbf{Beauty}. \textbf{Bold}: best per metric; \underline{underlined}: second-best.}
 \label{tab:per-beauty}
 \resizebox{0.7\linewidth}{!}{
 \begin{tabular}{l c c c c c c}
 \toprule
 \textbf{Method} & \textbf{R@1} & \textbf{R@5} & \textbf{R@10} & \textbf{N@1} & \textbf{N@5} & \textbf{N@10} \\
 \midrule
 \idlast    & 1.89 & 4.41 & 6.16 & 1.89 & 3.17 & 3.74 \\
 \semnn     & 1.69 & 3.95 & 5.32 & 1.69 & 2.85 & 3.30 \\
 \idsem     & 2.08 & 4.87 & 6.66 & 2.08 & 3.51 & 4.08 \\
 \midrule
 LightGCN   & \underline{2.18} & 5.21 & 7.21 & \underline{2.18} & \underline{3.71} & 4.36 \\
 SR-GNN     & 1.19 & 3.44 & 4.98 & 1.19 & 2.32 & 2.82 \\
 SASRec     & 1.21 & 3.99 & 6.21 & 1.21 & 2.59 & 3.31 \\
 HSTU       & 1.02 & 3.55 & 5.73 & 1.02 & 2.30 & 3.00 \\
 TIGER      & 1.48 & 4.40 & 6.41 & 1.48 & 2.95 & 3.60 \\
 LETTER & 1.22 & 3.87 & 6.11 & 1.22 & 2.53 & 3.26 \\
 CoFiRec    & 1.23 & 4.09 & 6.24 & 1.23 & 2.67 & 3.36 \\
 \midrule
 \oursone   & \textbf{2.87} & \textbf{5.79} & \underline{7.66} & \textbf{2.87} & \textbf{4.40} & \underline{5.01} \\
 \ourstwo   & \textbf{2.87} & \underline{5.78} & \textbf{7.85} & \textbf{2.87} & \textbf{4.40} & \textbf{5.07} \\
 \bottomrule
 \end{tabular}
 }
 \end{table}

  \begin{table}[htb]
 \centering
 \caption{Detailed results on \textbf{Sports}. \textbf{Bold}: best per metric; \underline{underlined}: second-best.}
 \label{tab:per-sports}
 \resizebox{0.7\linewidth}{!}{
 \begin{tabular}{l c c c c c c}
 \toprule
 \textbf{Method} & \textbf{R@1} & \textbf{R@5} & \textbf{R@10} & \textbf{N@1} & \textbf{N@5} & \textbf{N@10} \\
 \midrule
 \idlast    & 0.83 & 1.92 & 2.70 & 0.83 & 1.38 & 1.63 \\
 \semnn     & 0.69 & 1.85 & 2.68 & 0.69 & 1.28 & 1.55 \\
 \idsem     & 0.88 & 2.31 & 3.18 & 0.88 & 1.63 & 1.91 \\
 \midrule
 LightGCN   & 0.92 & 2.46 & 3.69 & 0.92 & 1.70 & 2.10 \\
 SR-GNN     & 0.43 & 1.50 & 2.61 & 0.43 & 0.97 & 1.33 \\
 SASRec     & 0.81 & 2.14 & 3.23 & 0.81 & 1.48 & 1.83 \\
 HSTU       & 0.42 & 1.41 & 2.48 & 0.42 & 0.90 & 1.25 \\
 TIGER      & 0.70 & 2.30 & 3.70 & 0.70 & 1.51 & 1.96 \\
 LETTER & 0.68 & 2.06 & 3.16 & 0.68 & 1.37 & 1.72 \\

 CoFiRec    & 0.69 & 2.45 & 3.82 & 0.69 & 1.58 & 2.02 \\
 \midrule
 \oursone   & \underline{1.58} & \underline{3.17} & \underline{4.27} & \underline{1.58} & \underline{2.41} & \underline{2.76} \\
 \ourstwo   & \textbf{1.59} & \textbf{3.19} & \textbf{4.66} & \textbf{1.59} & \textbf{2.42} & \textbf{2.90} \\
 \bottomrule
 \end{tabular}
 }
 \end{table}

  \begin{table}[htb]
 \centering
 \caption{Detailed results on \textbf{Toys}. \textbf{Bold}: best per metric; \underline{underlined}: second-best.}
 \label{tab:per-toys}
 \resizebox{0.7\linewidth}{!}{
 \begin{tabular}{l c c c c c c}
 \toprule
 \textbf{Method} & \textbf{R@1} & \textbf{R@5} & \textbf{R@10} & \textbf{N@1} & \textbf{N@5} & \textbf{N@10} \\
 \midrule
 \idlast    & 2.18 & 5.03 & 6.81 & 2.18 & 3.65 & 4.22 \\
 \semnn     & 2.53 & 6.14 & 8.04 & 2.53 & 4.40 & 5.01 \\
 \idsem     & 2.86 & 6.28 & 8.53 & 2.86 & 4.64 & 5.37 \\
 \midrule
 LightGCN   & 2.68 & 6.50 & 8.63 & 2.68 & 4.63 & 5.32 \\
 SR-GNN     & 1.21 & 2.75 & 3.77 & 1.21 & 1.99 & 2.32 \\
 SASRec     & 1.74 & 5.26 & 7.41 & 1.74 & 3.54 & 4.23 \\
 HSTU       & 1.40 & 4.38 & 6.40 & 1.40 & 2.91 & 3.56 \\
 TIGER      & 1.12 & 4.07 & 6.01 & 1.12 & 2.62 & 3.27 \\
 LETTER & 1.08 & 3.63 & 5.67 & 1.08 & 2.36 & 3.01 \\

 CoFiRec    & 0.84 & 3.59 & 5.63 & 0.84 & 2.22 & 2.89 \\
 \midrule
 \oursone   & \underline{3.71} & \textbf{7.04} & \underline{9.13} & \underline{3.71} & \underline{5.45} & \underline{6.12} \\
 \ourstwo   & \textbf{3.73} & \underline{7.01} & \textbf{9.44} & \textbf{3.73} & \textbf{5.46} & \textbf{6.25} \\
 \bottomrule
 \end{tabular}
 }
 \end{table}

 \begin{table}[htb]
 \centering
 \caption{Detailed results on \textbf{CDs}. \textbf{Bold}: best per metric; \underline{underlined}: second-best.}
 \label{tab:per-cds}
 \resizebox{0.7\linewidth}{!}{
 \begin{tabular}{l c c c c c c}
 \toprule
 \textbf{Method} & \textbf{R@1} & \textbf{R@5} & \textbf{R@10} & \textbf{N@1} & \textbf{N@5} & \textbf{N@10} \\
 \midrule
 \idlast    & 1.51 & 3.56 & 4.91 & 1.51 & 2.56 & 3.00 \\
 \semnn     & 0.41 & 0.86 & 1.20 & 0.41 & 0.64 & 0.75 \\
 \idsem     & 1.53 & \underline{3.63} & 5.02 & 1.53 & \underline{2.60} & 3.05 \\
 \midrule
 LightGCN   & 0.70 & 1.76 & 2.63 & 0.70 & 1.24 & 1.52 \\
 SR-GNN     & 0.14 & 0.42 & 0.63 & 0.14 & 0.28 & 0.35 \\
 SASRec     & 0.82 & 2.73 & 4.44 & 0.82 & 1.79 & 2.33 \\
 HSTU       & 0.09 & 3.48 & 5.62 & 0.09 & 2.21 & 2.91 \\
 TIGER & 1.06 & 3.45 & 5.06 & 1.06 & 2.26 & 2.78 \\
 LETTER & 1.12 & 3.47 & 5.14 & 1.12 & 2.31 & 2.85 \\
 CoFiRec    & 1.08 & 3.62 & 5.64 & 1.08 & 2.36 & 3.01 \\
 \midrule
 \oursone   & \underline{2.32} & \textbf{5.04} & \underline{6.21} & \underline{2.32} & \textbf{3.74} & \underline{4.13} \\
 \ourstwo   & \textbf{2.33} & \textbf{5.04} & \textbf{6.89} & \textbf{2.33} & \textbf{3.74} & \textbf{4.34} \\
 \bottomrule
 \end{tabular}
 }
 \end{table}

\FloatBarrier

\section{Detailed Results on More Datasets}
\label{sec:app:results-others}

For completeness, we further report the full per-metric results on the remaining ten benchmarks beyond the four Amazon Review datasets. These datasets cover diverse domains, including LastFM, Delicious, ML-1M, Yelp, MIND, Steam, GR-Children and GR-Comics, H\&M, and Amazon-M2-UK. We report Recall@1, Recall@5, Recall@10, NDCG@1, NDCG@5, and NDCG@10 in Tables~\ref{tab:per-lastfm}, \ref{tab:per-delicious}, \ref{tab:per-ml1m}, \ref{tab:per-yelp}, \ref{tab:per-mind}, \ref{tab:per-steam}, \ref{tab:per-gr-children}, \ref{tab:per-gr-comics}, \ref{tab:per-hm}, and \ref{tab:per-amazon-uk}, respectively.

Across these ten datasets, \ours achieves the best or second-best performance on a large fraction of metrics. These results show that the proposed heuristic remains competitive beyond the Amazon Review benchmarks and can generalize to a wide range of domains and dataset scales. At the same time, the results also reveal that \ours is not universally superior, suggesting that different datasets exhibit different degrees of shortcut solvability.

\begin{table}[htb]
 \centering
 \caption{Detailed results on \textbf{LastFM}. \textbf{Bold}: best per metric; \underline{underlined}: second-best.}
 \label{tab:per-lastfm}
 \resizebox{0.7\linewidth}{!}{
 \begin{tabular}{l c c c c c c}
 \toprule
 \textbf{Method} & \textbf{R@1} & \textbf{R@5} & \textbf{R@10} & \textbf{N@1} & \textbf{N@5} & \textbf{N@10} \\
 \midrule
 \idlast    & 1.10 & 2.11 & 3.58 & 1.10 & 1.63 & 2.11 \\
 \semnn     & 0.64 & \underline{3.39} & 5.14 & 0.64 & 2.09 & 2.66 \\
 \idsem     & 1.01 & \underline{3.39} & 4.59 & 1.01 & 2.22 & 2.60 \\
 \midrule
 LightGCN   & 0.64 & 2.57 & 4.59 & 0.64 & 1.57 & 2.22 \\
 SR-GNN     & 0.55 & 1.74 & 3.03 & 0.55 & 1.15 & 1.56 \\
 SASRec     & \underline{1.19} & \textbf{4.13} & \textbf{6.24} & \underline{1.19} & \textbf{2.65} & \textbf{3.32} \\
 HSTU       & 0.91 & 2.11 & 3.57 & 0.91 & 1.52 & 2.01 \\
 TIGER      & 0.18 & 1.56 & 3.39 & 0.18 & 0.85 & 1.45 \\
 LETTER & 0.55 & 2.11 & 4.04 & 0.55 & 1.35 & 1.97 \\
 \midrule
 \oursone   & \textbf{1.28} & \underline{3.39} & \underline{5.50} & \textbf{1.28} & \underline{2.38} & \underline{3.07} \\
 \ourstwo   & \textbf{1.28} & \underline{3.39} & 4.86 & \textbf{1.28} & \underline{2.38} & 2.87 \\
 \bottomrule
 \end{tabular}
 }
 \end{table}

 \begin{table}[htb]
 \centering
 \caption{Detailed results on \textbf{Delicious}. \textbf{Bold}: best per metric; \underline{underlined}: second-best.}
 \label{tab:per-delicious}
 \resizebox{0.7\linewidth}{!}{
 \begin{tabular}{l c c c c c c}
 \toprule
 \textbf{Method} & \textbf{R@1} & \textbf{R@5} & \textbf{R@10} & \textbf{N@1} & \textbf{N@5} & \textbf{N@10} \\
 \midrule
 \idlast    & 2.65 & 8.08 & \underline{11.56} & 2.65 & 5.52 & 6.64 \\
 \semnn     & 0.97 & 2.92 & 4.74 & 0.97 & 2.04 & 2.62 \\
 \idsem     & 2.79 & 8.22 & 11.42 & 2.79 & 5.64 & 6.68 \\
 \midrule
 LightGCN   & 1.67 & 5.71 & 7.80 & 1.67 & 3.70 & 4.37 \\
 SR-GNN     & 2.79 & 5.43 & 7.52 & 2.79 & 4.10 & 4.78 \\
 SASRec     & \textbf{4.18} & 8.64 & \underline{11.56} & \textbf{4.18} & \underline{6.59} & \textbf{7.57} \\
 HSTU       & 1.53 & 5.71 & 8.21 & 1.53 & 3.53 & 4.35 \\
 TIGER & \underline{3.76} & 5.71 & 6.82 & \underline{3.76} & 4.73 & 5.09 \\
 LETTER & 0.70 & 2.92 & 4.60 & 0.70 & 1.90 & 2.47 \\
 \midrule
 \oursone   & 3.62 & \underline{9.19} & 10.72 & 3.62 & 6.54 & 7.06 \\
 \ourstwo   & 3.62 & \textbf{9.47} & \textbf{12.12} & 3.62 & \textbf{6.69} & \underline{7.54} \\
 \bottomrule
 \end{tabular}
 }
 \end{table}

 \begin{table}[htb]
 \centering
 \caption{Detailed results on \textbf{ML-1M}. \textbf{Bold}: best per metric; \underline{underlined}: second-best.}
 \label{tab:per-ml1m}
 \resizebox{0.7\linewidth}{!}{
 \begin{tabular}{l c c c c c c}
 \toprule
 \textbf{Method} & \textbf{R@1} & \textbf{R@5} & \textbf{R@10} & \textbf{N@1} & \textbf{N@5} & \textbf{N@10} \\
 \midrule
 \idlast    & 2.76 & 8.34 & 12.27 & 2.76 & 5.60 & 6.86 \\
 \semnn     & 1.49 & 3.46 & 4.88 & 1.49 & 2.54 & 2.99 \\
 \idsem     & 2.90 & 8.69 & 12.62 & 2.90 & 5.83 & 7.09 \\
 \midrule
 LightGCN   & 1.52 & 4.97 & 8.39 & 1.52 & 3.22 & 4.32 \\
 SR-GNN     & 3.49 & 10.02 & 15.15 & 3.49 & 6.82 & 8.48 \\
 SASRec     & \textbf{5.68} & \textbf{17.25} & \textbf{25.43} & \textbf{5.68} & \textbf{11.58} & \textbf{14.21} \\
 HSTU       & 4.73 & 14.60 & 22.33 & 4.73 & 9.66 & 12.15 \\
 TIGER      & 3.56 & 12.30 & 18.54 & 3.56 & 7.97 & 9.98 \\
LETTER & \underline{5.43} & \underline{16.46} & \underline{23.73} & \underline{5.43} & \underline{10.95} & \underline{13.30} \\

 \midrule
 \oursone   & 4.52 & 12.12 & 14.82 & 4.52 & 8.34 & 9.25 \\
 \ourstwo   & 4.52 & 12.10 & 15.41 & 4.52 & 8.33 & 9.39 \\
 \bottomrule
 \end{tabular}
 }
 \end{table}

  \begin{table}[htb]
 \centering
 \caption{Detailed results on \textbf{Yelp}. \textbf{Bold}: best per metric; \underline{underlined}: second-best.}
 \label{tab:per-yelp}
 \resizebox{0.7\linewidth}{!}{
 \begin{tabular}{l c c c c c c}
 \toprule
 \textbf{Method} & \textbf{R@1} & \textbf{R@5} & \textbf{R@10} & \textbf{N@1} & \textbf{N@5} & \textbf{N@10} \\
 \midrule
 \idlast    & 0.32 & 1.06 & 1.86 & 0.32 & 0.69 & 0.94 \\
 \semnn     & 0.01 & 0.03 & 0.07 & 0.01 & 0.02 & 0.03 \\
 \idsem     & 0.24 & 0.75 & 1.23 & 0.24 & 0.50 & 0.65 \\
 \midrule
 LightGCN   & 0.12 & 0.47 & 0.78 & 0.12 & 0.30 & 0.40 \\
 SR-GNN     & \underline{0.42} & \underline{1.79} & \textbf{2.97} & \underline{0.42} & \underline{1.10} & \underline{1.48} \\
 SASRec     & 0.39 & 1.51 & 2.70 & 0.39 & 0.95 & 1.33 \\
 HSTU       & 0.29 & 1.29 & 2.21 & 0.29 & 0.78 & 1.08 \\
 TIGER      & 0.36 & 1.28 & 1.97 & 0.36 & 0.83 & 1.05 \\
 LETTER & \textbf{0.55} & \textbf{1.81} & \underline{2.90} & \textbf{0.55} & \textbf{1.19} & \textbf{1.54} \\
 \midrule
 \oursone   & 0.37 & 1.11 & 1.59 & 0.37 & 0.74 & 0.90 \\
 \ourstwo   & 0.37 & 1.12 & 1.87 & 0.37 & 0.74 & 0.98 \\
 \bottomrule
 \end{tabular}
 }
 \end{table}

  \begin{table}[htb]
 \centering
 \caption{Detailed results on \textbf{MIND}. \textbf{Bold}: best per metric; \underline{underlined}: second-best.}
 \label{tab:per-mind}
 \resizebox{0.7\linewidth}{!}{
 \begin{tabular}{l c c c c c c}
 \toprule
 \textbf{Method} & \textbf{R@1} & \textbf{R@5} & \textbf{R@10} & \textbf{N@1} & \textbf{N@5} & \textbf{N@10} \\
 \midrule
 \idlast    & 1.68 & 7.09 & 11.22 & 1.68 & 4.39 & 5.72 \\
 \semnn     & 0.07 & 0.14 & 0.25 & 0.07 & 0.10 & 0.14 \\
 \idsem     & 1.79 & 6.50 & 10.74 & 1.79 & 4.10 & 5.46 \\
 \midrule
 LightGCN   & 1.10 & 3.38 & 4.87 & 1.10 & 2.31 & 2.78 \\
 SR-GNN     & \textbf{7.29} & \textbf{17.00} & \underline{22.01} & \textbf{7.29} & \textbf{12.26} & \textbf{13.87} \\
 SASRec     & \underline{5.13} & \underline{15.22} & \textbf{22.73} & \underline{5.13} & \underline{10.25} & \underline{12.67} \\
 HSTU       & 2.45 & 9.01 & 14.50 & 2.45 & 5.72 & 7.49 \\
 TIGER & 3.93 & 11.81 & 17.72 & 3.93 & 7.88 & 9.79 \\
 LETTER & 4.09 & 12.35 & 19.25 & 4.09 & 8.24 & 10.46 \\
 \midrule
 \oursone   & 1.77 & 7.56 & 9.93 & 1.77 & 4.70 & 5.51 \\
 \ourstwo   & 1.77 & 7.57 & 9.50 & 1.77 & 4.70 & 5.32 \\
 \bottomrule
 \end{tabular}
 }
 \end{table}

\begin{table}[htb]
 \centering
 \caption{Detailed results on \textbf{STEAM}. \textbf{Bold}: best per metric; \underline{underlined}: second-best.}
 \label{tab:per-steam}
 \resizebox{0.7\linewidth}{!}{
 \begin{tabular}{l c c c c c c}
 \toprule
 \textbf{Method} & \textbf{R@1} & \textbf{R@5} & \textbf{R@10} & \textbf{N@1} & \textbf{N@5} & \textbf{N@10} \\
 \midrule
 \idlast    & 12.36 & 13.64 & 14.85 & 12.36 & 13.01 & 13.40 \\
 \semnn     & 12.36 & 13.88 & 14.60 & 12.36 & 13.16 & 13.40 \\
 \idsem     & 12.36 & 14.02 & 15.14 & 12.36 & 13.24 & 13.59 \\
 \midrule
 LightGCN   & 0.84 & 2.31 & 3.55 & 0.84 & 1.58 & 1.98 \\
 SR-GNN     & 12.15 & 15.80 & \underline{18.71} & 12.15 & 14.00 & 14.93 \\
 SASRec     & 1.43 & 5.32 & 8.81 & 1.43 & 3.37 & 4.49 \\
 HSTU       & 1.41 & 5.17 & 8.41 & 1.41 & 3.28 & 4.32 \\
 TIGER      & 11.75 & 15.08 & 17.60 & 11.75 & 13.43 & 14.24 \\
 LETTER & \textbf{12.95} & \textbf{17.69} & \textbf{20.98} & \textbf{12.95} & \textbf{15.36} & \textbf{16.42} \\
 \midrule
 \oursone   & \underline{12.79} & \underline{16.00} & 17.23 & \underline{12.79} & \underline{14.43} & 14.85 \\
 \ourstwo   & \underline{12.79} & \underline{16.00} & 17.65 & \underline{12.79} & \underline{14.43} & \underline{14.95} \\
 \bottomrule
 \end{tabular}
 }
 \end{table}

  \begin{table}[htb]
 \centering
 \caption{Detailed results on \textbf{GR-Children}. \textbf{Bold}: best per metric; \underline{underlined}: second-best.}
 \label{tab:per-gr-children}
 \resizebox{0.7\linewidth}{!}{
 \begin{tabular}{l c c c c c c}
 \toprule
 \textbf{Method} & \textbf{R@1} & \textbf{R@5} & \textbf{R@10} & \textbf{N@1} & \textbf{N@5} & \textbf{N@10} \\
 \midrule
 \idlast    & 4.87 & 11.01 & 15.31 & 4.87 & 8.01 & 9.39 \\
 \semnn     & 0.72 & 2.19 & 3.56 & 0.72 & 1.47 & 1.91 \\
 \idsem     & 4.04 & 8.30 & 10.68 & 4.04 & 6.23 & 7.00 \\
 \midrule
 LightGCN   & 4.17 & 9.74 & 12.73 & 4.17 & 7.07 & 8.03 \\
 SR-GNN     & 4.82 & 11.03 & 15.29 & 4.82 & 8.00 & 9.37 \\
 SASRec     & 5.30 & 14.74 & \underline{20.83} & 5.30 & 10.14 & 12.10 \\
 HSTU       & 5.02 & 13.81 & 20.06 & 5.02 & 9.49 & 11.50 \\
 TIGER      & 5.42 & 12.38 & 17.11 & 5.42 & 8.96 & 10.48 \\
 LETTER & \textbf{8.66} & \textbf{19.10} & \textbf{25.76} & \textbf{8.66} & \textbf{13.99} & \textbf{16.14} \\
 \midrule
 \oursone   & \underline{7.14} & \underline{15.29} & 17.89 & \underline{7.14} & \underline{11.34} & 12.23 \\
 \ourstwo   & \underline{7.14} & \underline{15.29} & 18.80 & \underline{7.14} & \underline{11.34} & \underline{12.46} \\
 \bottomrule
 \end{tabular}
 }
 \end{table}

  \begin{table}[htb]
 \centering
 \caption{Detailed results on \textbf{GR-Comics}. \textbf{Bold}: best per metric; \underline{underlined}: second-best.}
 \label{tab:per-gr-comics}
 \resizebox{0.7\linewidth}{!}{
 \begin{tabular}{l c c c c c c}
 \toprule
 \textbf{Method} & \textbf{R@1} & \textbf{R@5} & \textbf{R@10} & \textbf{N@1} & \textbf{N@5} & \textbf{N@10} \\
 \midrule
 \idlast    & 14.24 & 20.76 & 23.62 & 14.24 & 17.77 & 18.70 \\
 \semnn     & 4.47 & 10.57 & 12.86 & 4.47 & 7.56 & 8.30 \\
 \idsem     & 13.72 & 20.74 & 23.74 & 13.72 & 17.46 & 18.43 \\
 \midrule
 LightGCN   & 14.49 & 22.30 & 25.36 & 14.49 & 18.63 & 19.62 \\
 SR-GNN     & 6.05 & 10.60 & 13.26 & 6.05 & 8.40 & 9.26 \\
 SASRec     & 8.62 & 19.31 & 25.33 & 8.62 & 14.14 & 16.08 \\
 HSTU       & 8.96 & 20.02 & 25.87 & 8.96 & 14.68 & 16.57 \\
 TIGER      & 11.68 & 17.96 & 21.18 & 11.68 & 14.97 & 16.02 \\
 LETTER & \textbf{17.97} & \textbf{27.66} & \textbf{32.58} & \textbf{17.97} & \textbf{23.05} & \textbf{24.64} \\
 \midrule
 \oursone   & \underline{17.79} & \underline{26.41} & 28.69 & \underline{17.79} & \underline{22.35} & 23.12 \\
 \ourstwo   & \underline{17.79} & \underline{26.41} & \underline{30.48} & \underline{17.79} & \underline{22.35} & \underline{23.66} \\
 \bottomrule
 \end{tabular}
 }
 \end{table}

 \begin{table}[htb]
 \centering
 \caption{Detailed results on \textbf{H\&M}. \textbf{Bold}: best per metric; \underline{underlined}: second-best.}
 \label{tab:per-hm}
 \resizebox{0.7\linewidth}{!}{
 \begin{tabular}{l c c c c c c}
 \toprule
 \textbf{Method} & \textbf{R@1} & \textbf{R@5} & \textbf{R@10} & \textbf{N@1} & \textbf{N@5} & \textbf{N@10} \\
 \midrule
 \idlast    & 0.29 & 0.91 & 1.49 & 0.29 & 0.60 & 0.79 \\
 \semnn     & 1.58 & 4.15 & 5.46 & 1.58 & 2.91 & 3.33 \\
 \idsem     & 1.99 & 4.63 & 5.88 & 1.99 & 3.37 & 3.77 \\
 \midrule
 LightGCN   & 1.61 & 3.59 & 4.45 & 1.61 & 2.66 & 2.93 \\
 SR-GNN     & 1.68 & 4.21 & 5.99 & 1.68 & 2.96 & 3.54 \\
 SASRec     & 1.36 & 4.42 & 6.87 & 1.36 & 2.90 & 3.68 \\
 HSTU       & 1.27 & 4.12 & 6.45 & 1.27 & 2.70 & 3.45 \\
 TIGER      & 1.60 & 5.12 & 7.30 & 1.60 & 3.40 & 4.10 \\
 LETTER & \underline{3.50} & \underline{7.19} & 9.55 & \underline{3.50} & \underline{5.41} & 6.17 \\
 \midrule
 \oursone   & \textbf{4.92} & \textbf{9.97} & \underline{11.46} & \textbf{4.92} & \textbf{7.56} & \underline{8.06} \\
 \ourstwo   & \textbf{4.92} & \textbf{9.97} & \textbf{13.51} & \textbf{4.92} & \textbf{7.56} & \textbf{8.70} \\
 \bottomrule
 \end{tabular}
 }
 \end{table}

 \begin{table}[htb]
 \centering
 \caption{Detailed results on \textbf{Amazon-UK}. \textbf{Bold}: best per metric; \underline{underlined}: second-best.}
 \label{tab:per-amazon-uk}
 \resizebox{0.7\linewidth}{!}{
 \begin{tabular}{l c c c c c c}
 \toprule
 \textbf{Method} & \textbf{R@1} & \textbf{R@5} & \textbf{R@10} & \textbf{N@1} & \textbf{N@5} & \textbf{N@10} \\
 \midrule
 \idlast    & 17.79 & 29.18 & 33.97 & 17.79 & 23.83 & 25.38 \\
 \semnn     & 13.50 & 26.30 & 30.69 & 13.50 & 20.33 & 21.76 \\
 \idsem     & \underline{18.16} & 30.27 & 35.07 & \underline{18.16} & \underline{24.61} & \underline{26.16} \\
 \midrule
 LightGCN   & \textbf{19.29} & \textbf{33.67} & \textbf{39.31} & \textbf{19.29} & \textbf{26.95} & \textbf{28.78} \\
 SR-GNN     & 4.03 & 11.77 & 15.28 & 4.03 & 7.98 & 9.12 \\
 SASRec     & 6.84 & 17.25 & 23.10 & 6.84 & 12.20 & 14.10 \\
 HSTU       & 12.14 & 22.28 & 27.39 & 12.14 & 17.45 & 19.10 \\
 TIGER      & 3.76 & 10.34 & 13.39 & 3.76 & 7.11 & 8.10 \\
 LETTER & 2.94 & 9.74 & 14.32 & 2.94 & 6.31 & 7.78 \\
 \midrule
 \oursone   & 15.21 & 30.01 & 35.63 & 15.21 & 23.04 & 24.87 \\
 \ourstwo   & 15.21 & \underline{30.28} & \underline{36.36} & 15.21 & 23.16 & 25.13 \\
 \bottomrule
 \end{tabular}
 }
 \end{table}

 \FloatBarrier

 \section{Limitations}
\label{sec:limitation}
 Our study has several limitations. First, the three shortcut structures we identify are not intended to be an exhaustive explanation of benchmark behavior. Sequential recommendation datasets may contain other shortcut signals, such as popularity effects, temporal regularities, repeated consumption patterns, or preprocessing artifacts. Our diagnostics provide a useful lens for interpreting the behavior of \ours, but they do not fully characterize all factors that influence model performance. Second, our analysis is based on a finite set of datasets and baselines. Although we evaluate on 14 datasets covering different domains, sequence lengths, graph structures, and feature signals, they still cannot represent all sequential recommendation scenarios. Similarly, while we include representative traditional, graph-based, sequential, and generative recommenders, new architectures or stronger implementations may behave differently. Third, our heuristic uses fixed hyperparameters and fixed retrieval budgets across datasets. This design keeps the heuristic simple and avoids dataset-specific configurations, but it may not be optimal for every dataset. Finally, our goal is not to argue that advanced sequential or generative recommenders are unnecessary. Instead, we show that some widely used benchmarks may not be sufficient to demonstrate their benefits. Future work can extend our diagnostics to more datasets, richer evaluation protocols, and more settings to better understand when advanced recommendation models provide genuine advantages.

\section{Broader Impacts}
\label{app:sec:impact}

This work studies evaluation practice in sequential recommendation. Its main positive impact is to encourage more reliable and capability-aware benchmarking. By showing that some widely used datasets can be solved by simple shortcut signals, our analysis may help researchers avoid overclaiming model capabilities based on narrow benchmark results. It may also help benchmark creators release more transparent datasets by reporting diagnostic properties such as transition branching, feature-smoothness, and history dependence.

More careful benchmark selection can benefit both the research community and downstream users of recommender systems. If models are evaluated on datasets that better match their claimed capabilities, progress in semantic modeling, generative retrieval, and long-context recommendation can be measured more accurately. This may reduce wasted effort on optimizing for shortcut-solvable benchmarks and encourage the development of models that address harder and more realistic recommendation challenges.

Overall, we hope this work promotes more transparent evaluation practice in sequential recommendation, where both model designers and benchmark creators analyze dataset properties before drawing broad conclusions about model capability.


\newpage
\section*{NeurIPS Paper Checklist}

\begin{enumerate}

\item {\bf Claims}
    \item[] Question: Do the main claims made in the abstract and introduction accurately reflect the paper's contributions and scope?
    \item[] Answer: \answerYes{} 
    \item[] Justification: The abstract and introduction contain the paper’s main contributions and scope.

\item {\bf Limitations}
    \item[] Question: Does the paper discuss the limitations of the work performed by the authors?
    \item[] Answer: \answerYes{} 
    \item[] Justification: We include a Limitations Section in Appendix ~\ref{sec:limitation}.

\item {\bf Theory assumptions and proofs}
    \item[] Question: For each theoretical result, does the paper provide the full set of assumptions and a complete (and correct) proof?
    \item[] Answer: \answerNA{} 
    \item[] Justification: This paper does not include theoretical result.

    \item {\bf Experimental result reproducibility}
    \item[] Question: Does the paper fully disclose all the information needed to reproduce the main experimental results of the paper to the extent that it affects the main claims and/or conclusions of the paper (regardless of whether the code and data are provided or not)?
    \item[] Answer: \answerYes{} 
    \item[] Justification: We provide detailed experimental setup in Section~\ref{sec:problem}, Appendix~\ref{app:sec:datasets}, Appendix~\ref{app:sec:baselines} and Appendix~\ref{app:sec:implementation_details}.

\item {\bf Open access to data and code}
    \item[] Question: Does the paper provide open access to the data and code, with sufficient instructions to faithfully reproduce the main experimental results, as described in supplemental material?
    \item[] Answer: \answerYes{}{} 
    \item[] Justification: We provide the code, dataset files or link used in experiments: https://github.com/haoyuhan1/GraphRec/.

\item {\bf Experimental setting/details}
    \item[] Question: Does the paper specify all the training and test details (e.g., data splits, hyperparameters, how they were chosen, type of optimizer) necessary to understand the results?
    \item[] Answer: \answerYes{} 
    \item[] Justification: The full details are provided in Section~\ref{sec:problem}, Appendix~\ref{app:sec:implementation_details} and code.

\item {\bf Experiment statistical significance}
    \item[] Question: Does the paper report error bars suitably and correctly defined or other appropriate information about the statistical significance of the experiments?
    \item[] Answer: \answerYes{} 
    \item[] Justification: The used heuristic is fixed, therefore the variance is 0.

\item {\bf Experiments compute resources}
    \item[] Question: For each experiment, does the paper provide sufficient information on the computer resources (type of compute workers, memory, time of execution) needed to reproduce the experiments?
    \item[] Answer: \answerYes{} 
    \item[] Justification: The paper provides details in Appendix~\ref{app:sec:implementation_details}.
    
\item {\bf Code of ethics}
    \item[] Question: Does the research conducted in the paper conform, in every respect, with the NeurIPS Code of Ethics \url{https://neurips.cc/public/EthicsGuidelines}?
    \item[] Answer: \answerYes{} 
    \item[] Justification: The research conducted in the paper conforms, in every respect, with the NeurIPS Code of Ethics.

\item {\bf Broader impacts}
    \item[] Question: Does the paper discuss both potential positive societal impacts and negative societal impacts of the work performed?
    \item[] Answer: \answerYes{} 
    \item[] Justification: The paper includes a Broader Impacts section in Appendix~\ref{app:sec:impact}.
    
\item {\bf Safeguards}
    \item[] Question: Does the paper describe safeguards that have been put in place for responsible release of data or models that have a high risk for misuse (e.g., pre-trained language models, image generators, or scraped datasets)?
    \item[] Answer: \answerNA{} 
    \item[] Justification: The released code does not pose a high risk of misuse.

\item {\bf Licenses for existing assets}
    \item[] Question: Are the creators or original owners of assets (e.g., code, data, models), used in the paper, properly credited and are the license and terms of use explicitly mentioned and properly respected?
    \item[] Answer: \answerYes{} 
    \item[] Justification: We provide details in Appendices~\ref{app:sec:datasets} and ~\ref{app:sec:implementation_details}.

\item {\bf New assets}
    \item[] Question: Are new assets introduced in the paper well documented and is the documentation provided alongside the assets?
    \item[] Answer: \answerYes{} 
    \item[] Justification: We provide a detailed README file for the released code.

\item {\bf Crowdsourcing and research with human subjects}
    \item[] Question: For crowdsourcing experiments and research with human subjects, does the paper include the full text of instructions given to participants and screenshots, if applicable, as well as details about compensation (if any)? 
    \item[] Answer: \answerNA{} 
    \item[] Justification: The paper involves neither crowdsourcing nor research with human subjects.

\item {\bf Institutional review board (IRB) approvals or equivalent for research with human subjects}
    \item[] Question: Does the paper describe potential risks incurred by study participants, whether such risks were disclosed to the subjects, and whether Institutional Review Board (IRB) approvals (or an equivalent approval/review based on the requirements of your country or institution) were obtained?
    \item[] Answer: \answerNA{} 
    \item[] Justification: The paper does not involve crowdsourcing nor research with human subjects

\item {\bf Declaration of LLM usage}
    \item[] Question: Does the paper describe the usage of LLMs if it is an important, original, or non-standard component of the core methods in this research? Note that if the LLM is used only for writing, editing, or formatting purposes and does \emph{not} impact the core methodology, scientific rigor, or originality of the research, declaration is not required.
    \item[] Answer: \answerNA{} 
    \item[] Justification: The LLM is only used for formatting and wording.

\end{enumerate}

\end{document}